\documentclass[]{spie}  
\usepackage[]{graphicx}
\usepackage{amsmath}
\usepackage{amsfonts}
\usepackage{amssymb}
\usepackage{subfigure}

\usepackage[latin1]{inputenc}  
\usepackage[english]{babel}

\newcommand{\ud}{\, \mathrm{d}}

\title{Stellar intensity interferometry:\\Optimizing air Cherenkov telescope array layouts} 

\author{Hannes Jensen\supit{a}, Dainis Dravins\supit{a}, Stephan LeBohec\supit{b}, Paul D. Nuñez\supit{b}
\skiplinehalf
\supit{a}Lund Observatory, Box 43, SE-22100 Lund, Sweden \\
\supit{b}Department of Physics and Astronomy, The University of Utah, 115 South 1400 East,\\ Salt Lake City, UT 84112-0830, U.S.A.\\
}

\authorinfo{\supit{a} hannesj1@gmail.com, dainis@astro.lu.se, www.astro.lu.se}

\begin{document} 
  \maketitle 

\begin{abstract}
Kilometric-scale optical imagers seem feasible to realize by intensity interferometry, using telescopes primarily erected for measuring Cherenkov light induced by gamma rays.  Planned arrays envision 50--100 telescopes, distributed over some 1--4 km$^2$.  Although array layouts and telescope sizes will primarily be chosen for gamma-ray observations, also their interferometric performance may be optimized.  Observations of stellar objects were numerically simulated for different array geometries, yielding signal-to-noise ratios for different Fourier components of the source images in the interferometric $(u,v)$-plane.  Simulations were made for layouts actually proposed for future Cherenkov telescope arrays, and for subsets with only a fraction of the telescopes.  All large arrays provide dense sampling of the $(u,v)$-plane due to the sheer number of telescopes, irrespective of their geographic orientation or stellar coordinates.  However, for improved coverage of the $(u,v)$-plane and a wider variety of baselines (enabling better image reconstruction), an exact east-west grid should be avoided for the numerous smaller telescopes, and repetitive geometric patterns avoided for the few large ones.  Sparse arrays become severely limited by a lack of short baselines, and to cover astrophysically relevant dimensions between 0.1--3 milliarcseconds in visible wavelengths, baselines between pairs of telescopes should cover the whole interval 30--2000 m.
 
\end{abstract}


\section{IMAGING STARS WITH INTENSITY INTERFEROMETRY}

\subsection{Introduction}

A small number of the apparently largest stars have now been imaged with phase/amplitude interferometers, revealing their oblate shapes caused by rapid rotation or perhaps being surrounded by circumstellar winds, disks, or shells.  Clearly, improving the angular resolution by just another order of magnitude would enable many more phenomena to be resolved\cite{dravins10}.  However, atmospheric turbulence makes the required interferometry over hundreds of meters quite challenging for phase/amplitude interferometers which require optical path-length precisions within a fraction of an optical wavelength.

Optical intensity interferometry currently seems the most realistic way to realize imaging of stars on scales well below one milliarcsecond (mas)\cite{lebohec06,lebohec08}. Using a simple $\lambda/D$ criterion, 1 mas resolution at $\lambda$=500 nm requires a baseline around 100 meters, while 1 km enables 100 $\mu$as. For the forthcoming large arrays of Cherenkov telescopes, extensions over some 2 km are discussed, and if such could be utilized at $\lambda$=350 nm, resolutions could approach 30 $\mu$as.

\subsection{Theory of Intensity Interferometry}

Intensity interferometry measures correlations in light intensity fluctuations between pairs of telescopes\cite{hanburybrown74,labeyrie06}.  This correlation is related to the squared modulus of the complex degree of coherence of the light:
\begin{equation}
\label{intcorr}
|\gamma|^2 = \frac{\langle \Delta I_1 \Delta I_2 \rangle}{\langle I_1 \rangle \langle I_2 \rangle}
\end{equation}
where $I_i$ and $\Delta I_i$ are the intensity and intensity fluctuations respectively, measured by telescope $i$ and $\langle \rangle$ denotes the time average.

The complex coherence is proportional to the Fourier transform of the stellar surface brightness, $\Theta$, sampled at a point $(u,v)$ in the Fourier plane, according to the van Cittert-Zernike theorem:
\begin{equation}
\gamma(u,v) \propto \iint \Theta(l,m) e^{-2 \pi i (ul+vm)} \ud l \ud m
\end{equation}
where $l,m$ are (angular) positions in the sky. So, in short, each pair of telescopes in an intensity interferometer measures the square of the magnitude of the Fourier transform of the target at one point in the Fourier plane. 

Long telescope separations, or baselines, measure high-frequency components, corresponding to small structures on the target, while short baselines sample the low frequencies. For a baseline $\mathbf{B} = (B_{\mathrm{North}}, B_{\mathrm{East}})$ the associated Fourier-plane coordinates are $(u,v) = \frac{1}{\lambda}(B_{\mathrm{North}}, B_{\mathrm{East}})$.

For an array of $N$ telescopes there are $N(N-1)/2$ (not necessarily unique) baselines. As the target of observation moves across the sky, the projected baselines, $\mathbf{B}_p$, will change, with each telescope pair tracing out an ellipse in the Fourier plane according to the following expression\cite{segransan07}:
\begin{small}
\begin{equation}
\label{rotationsynth}
(u,v,w) = 
\frac{1}{\lambda} \mathbf{B}_p =
\frac{1}{\lambda} \begin{pmatrix}
-\sin l \sin h & \cos h & \cos l \sin h\\
\sin l \cos h \sin \delta + \cos l \cos \delta & \sin h \sin \delta & -\cos l \cos h \sin \delta + \sin l \cos \delta \\
-\sin l \cos h \cos \delta + \cos l \sin \delta & -\sin h \cos \delta & \cos l \cos h \cos \delta + \sin l \sin \delta
\end{pmatrix}
\begin{pmatrix}
B_{\mathrm{North}} \\
B_{\mathrm{East}}\\
0
\end{pmatrix}
\end{equation}
\end{small}

where $l$ is the latitude of the telescope array and $\delta$ and $h$ are the declination and hour angle of the star. The $w$ component corresponds to the time delay between the two telescopes. This effect, known as \emph{Earth rotation synthesis}, greatly increases the Fourier-plane coverage of a telescope array.

From the Fourier magnitude, it is possible to both reconstruct the image of the target\cite{nunez10} (provided the sampling is good enough), or fit a model to the data.

\begin{figure}[h!]
\centering
\includegraphics[width=12cm]{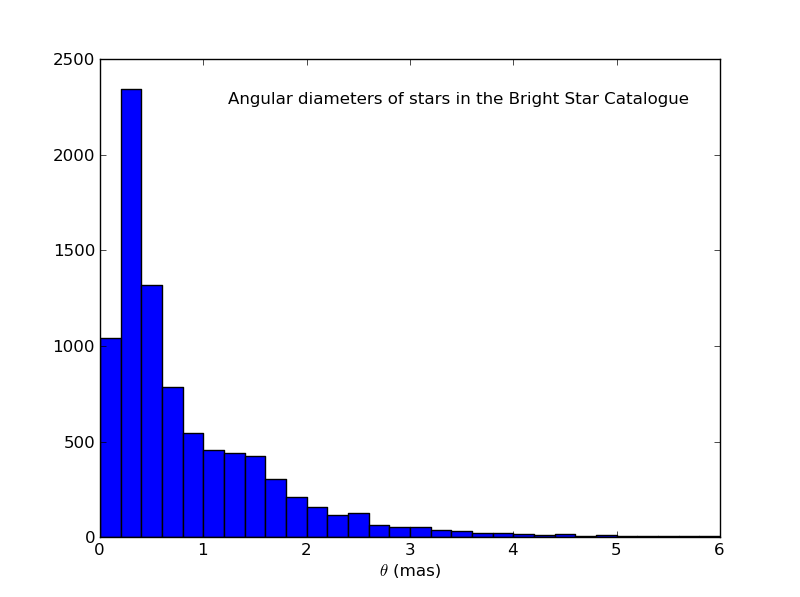} 
\caption{Approximated angular diameters of the stars in the Bright Star Catalogue (BSC)\cite{hoffleit95}, containing all stars in the sky with $m_V<6.5$ (a few outliers have been omitted for clarity). Effective temperatures were estimated from their $B-V$ color index, and then used to calculate the angular diameters.}
\label{theta_hist}
\end{figure}

Optimizing the telescope placement in an interferometric array is a complex problem that has been widely studied for optical and radio interferometers (e.g., Refs.\ \citenum{keto97,millour08,boone01}).  The optimal configuration depends upon the task: if the objective is to study very small structures, long baselines are required. If, on the other hand, one wishes to measure large-scale structures, shorter baselines are necessary. For a general-purpose array, it is usually preferable to have a healthy mixture of short and long baselines, sampling the $(u,v)$-plane as uniformly as possible.  In some interferometers this is satisfied by movable telescopes; in others one maintains similar diffraction-limited resolutions for observations at different wavelengths, placing telescopes in self-similar patterns (e.g., logarithmic spirals), while still others may be constrained by the local geography.  In any case it is almost always best to avoid regularly spaced telescopes, since such placement results in duplicate baselines.

\subsection{Astrophysical Targets}

Ideal targets for intensity interferometry are hot and bright stars\cite{dravins10}. The limiting magnitude depends on the size of the flux collectors, the detectors with their electronics, and the integration time, but simulations and previous experience suggest a limiting magnitude around $m_V \sim 9$\cite{lebohec06}, while initial observing programs will likely be limited to $m_V \lesssim 6$.

The Bright Star Catalogue (BSC)\cite{hoffleit95} lists all stars brighter than $m_V = 6.5$ (a total of 9096), and can be used to illustrate the parameters of likely observational targets. The $B-V$ colors of the BSC entries were converted to theoretical effective temperatures by fitting a polynomial to data relating $T_{eff}$ and $B-V$ color index from Bessel et al.\cite{bessel98}.  These effective temperatures were then used to calculate the angular diameters assuming the stars are blackbodies with uniform circular disks; see Fig.\ \ref{theta_hist}. 

Most bright stars have angular diameters on the order of one milliarcsecond (mas) or less. Such a star is resolved by  telescopic baselines of $\sim$100 m, so an array must provide baselines at least this long to be able to observe any significant number of objects.  However, it is important to note that short baselines are also crucial. An array lacking baselines shorter than $\sim 100$ m will have too narrow a field of view to image the shapes of most stars.

\subsection{Technical Requirements}

Intensity interferometry places substantial technical requirements on the detection equipment used (for a discussion of its technical aspects, see Ref.\ \citenum{lebohec10}).  Very high time-resolution detectors are required to measure the rapid intensity fluctuations, and fast hardware is needed for calculating correlation functions. However, by far the most expensive components are the flux collectors.  The technique is sensitive to the square of the received light intensity, and large amounts of starlight need to be collected. On the other hand, there is no need for high-quality optics -- a time resolution on the order of 1 ns, say, corresponds to a light travel distance of 30 cm, and optical imperfections comparable to such a magnitude can be tolerated.  Despite the tolerance for optical imperfections, the cost associated with a large array of $\sim$10 m telescopes is substantial, so a very attractive possibility is to operate intensity interferometry as an auxiliary program on arrays set up for other purposes.

Seemingly ideal for this purpose are those air Cherenkov telescopes that are being erected for gamma-ray astronomy (e.g., Refs. \citenum{hinton09,voelk09}).  These measure the feeble Cherenkov light in air produced by cascades of secondary particles initiated by highly energetic gamma rays.  They must have a time resolution of a few nanoseconds (set by the duration of the Cherenkov light flash); they must be sensitive to short optical wavelengths (Cherenkov light is bluish); they must be large (Cherenkov light is faint), and they need to be separated by distances of $\sim$50~m to $\sim$200~m (extent of the Cherenkov light-pool onto the ground). 

\section{EXISTING CHERENKOV TELESCOPE ARRAYS}

\begin{figure}[tbh]
\centering
\includegraphics[width=14cm]{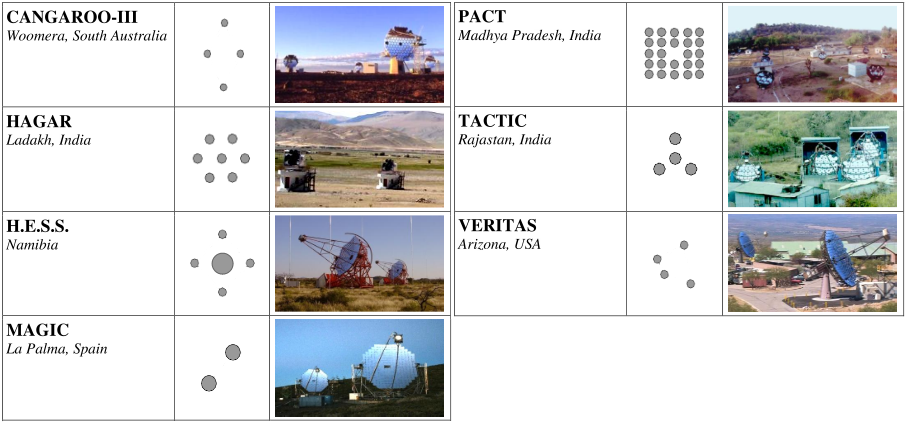} 
\caption{Overview of existing air Cherenkov telescope arrays. The geometric layout is shown with north upwards (the scale varies between different arrays in the figure; see Table \ref{existing_arrays_table} for array sizes).}
\label{cherenkov_arrays}
\end{figure}

A number of air Cherenkov telescope facilities are already in operation, with the major ones summarized in Fig.~\ref{cherenkov_arrays} and Table \ref{existing_arrays_table}. 

\begin{table}[h]
\caption{Properties of some existing Cherenkov arrays. $N$ denotes the number of telescopes, $A$ is the light collection area of each telescope, $n$ is the number of unique baselines simultaneously available, $B_{min}, B_{max}$ indicates the range of baselines for observation at zenith. The corresponding range of angular diameters in milliarcseconds $(1.22 \lambda/b)$ for observations at 400 nm is indicated by  $\theta_{min}, \theta_{max}$.}
\begin{center}
\begin{tabular}{rccccc|}
Array & N & A (m$^2$) & $n$ & $B_{min}, B_{max}$ (m) & $\theta_{min}, \theta_{max}$ (mas)\\
\hline \hline
CANGAROO-III\cite{kubo04} &	4 & 	57 & 		6 & 	100, 184 & 	0.5, 1.0 \\
HAGAR\cite{chitnis05} & 		7 & 	7 & 		21 &	50, 100	&	1.0, 2.0 \\
H.E.S.S.-I\cite{hinton2004} &	4 & 	108 & 		6 & 	120, 170 & 	0.6,  0.8	\\
MAGIC\cite{cortina05} & 		2 & 	227 & 		1 & 	85, 85 & 	1.2, 1.2 \\
PACT\cite{bhat00} &			25 &	9.5 &		300 &	20, 128 &	0.8, 5.0 \\
TACTIC\cite{koul07} &		4 &		7 &			6 &		17, 20 &	5.0, 6.0 \\
VERITAS\cite{holder06} &		4& 		113 & 		6 & 	80, 140 & 	0.7, 1.2	\\
\end{tabular}
\end{center}
\label{existing_arrays_table}
\end{table}

\subsection{Interferometric capabilities of existing arrays}

Already the baselines offered by some of the existing arrays may be adequate for starting intensity interferometry experiments.  
Their adequately large flux collectors should provice excellent signal-to-noise ratios, but the longest baselines do not reach more than some 200 meters and their achievable angular resolution essentially overlaps with that feasible with existing phase interferometers (though observations could be made in the blue or violet, where the contrast of many stellar features is expected to be higher).  Table \ref{existing_arrays_table} lists some of their relevant properties.

\begin{figure}[tbh]
\centering
\includegraphics[width=12cm]{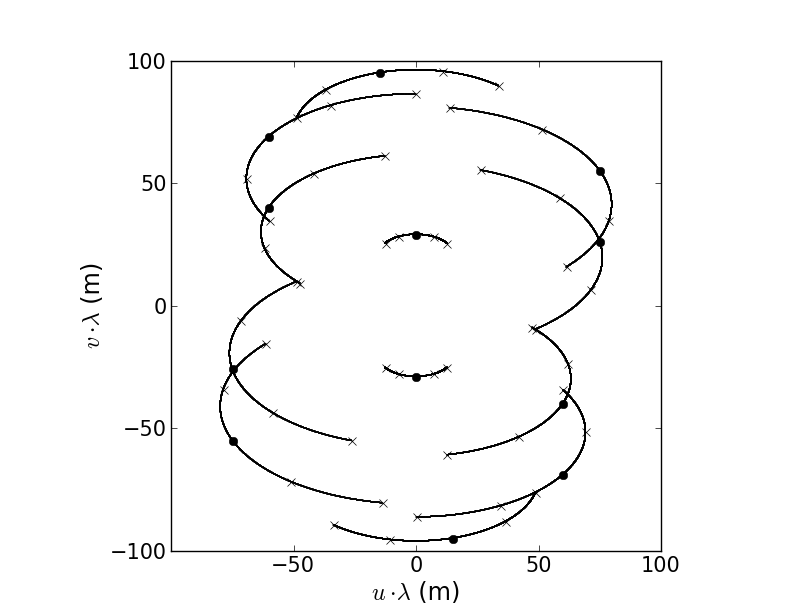} 
\caption{Simulated $(u,v)$-coverage of the VERITAS array (Fig.\ \ref{cherenkov_arrays} and Table \ref{existing_arrays_table}) for an 8-hour observation of a star moving through zenith. Each pair of telescopes traces out two partial ellipses in the Fourier plane, since the Fourier magnitude in a point $(u,v)$ will be the same as in the point $(-u,-v)$. The filled circles indicate the snapshot Fourier plane coverage when the star is in the zenith. The small crosses mark 2-hour intervals before and after the meridian passage.}
\label{veritas_uv}
\end{figure}

Figure \ref{veritas_uv} shows a simulation of the Fourier plane coverage of the baselines possible with the VERITAS array for an 8-hour observation of a star moving through zenith, cf.\ Eq.\ \eqref{rotationsynth}. While the $(u,v)$-coverage is increased greatly through Earth rotation, it is still rather sparse and does not extend very far.

\section{FUTURE CHERENKOV TELESCOPE ARRAYS}
The largest current projects for future observatories are CTA (Cherenkov Telescope Array)\cite{ctawebsite}, and AGIS (Advanced Gamma-ray Imaging System)\cite{agiswebsite}. Either of them is envisioned to comprise around 50-100 flux collectors with diameters up to $\sim$20~m spread over kilometer-scale distances.  These would be excellent facilities for intensity interferometry, both in terms of resolution and Fourier-plane sampling, and their design studies now also incorporate their possible use for optical interferometrry. For such use, appropriate data analysis software would digitally synthesize baselines between all possible pairs of telescopes \cite{lebohec06,lebohec08}.

\subsection{Candidate configurations}

Any dedicated interferometer can be optimized for the best coverage of the interferometric $(u,v)$-plane.  However, for Cherenkov telescope arrays, intensity interferometry is a secondary application and telescope placements and sizes will primarily be chosen for their gamma-ray observations.  Nevertheless, also the interferometric performance may be optimized, without compromising the gamma-ray performance nor adding extra costs.

\begin{table}[b]
\caption{Properties of the three examined candidate configurations. $N$ is the number of telescopes, $A$ is the light collection area of each type of telescope, $n$ is the number of unique baselines available, $B_{min}, B_{max}$ indicates the range of baselines for observations in zenith. The corresponding range of angular diameters in milliarcseconds $(1.22 \lambda/b)$ for observations at 400~nm is indicated by  $\theta_{min}, \theta_{max}$.}
\begin{center}
\begin{tabular}{lrrrrr|}
Array & N & A (m$^2$) & $n$ & $B_{min}, B_{max}$ (m) & $\theta_{min}, \theta_{max}$ (mas)\\
\hline \hline
Conf. B & 42 & 113, 415 & 253 & 32, 759 & 0.13, 3.2 \\
Conf. D & 57 & 113 & 487 & 170, 2180 & 0.05, 0.6 \\
Conf. I & 77 & 28, 113, 415 & 1606 & 90, 2200 & 0.05,  1.13	
\end{tabular}
\end{center}
\label{configurationstable}
\end{table}

\begin{figure}[tbh]
\centering
\includegraphics[width=14cm]{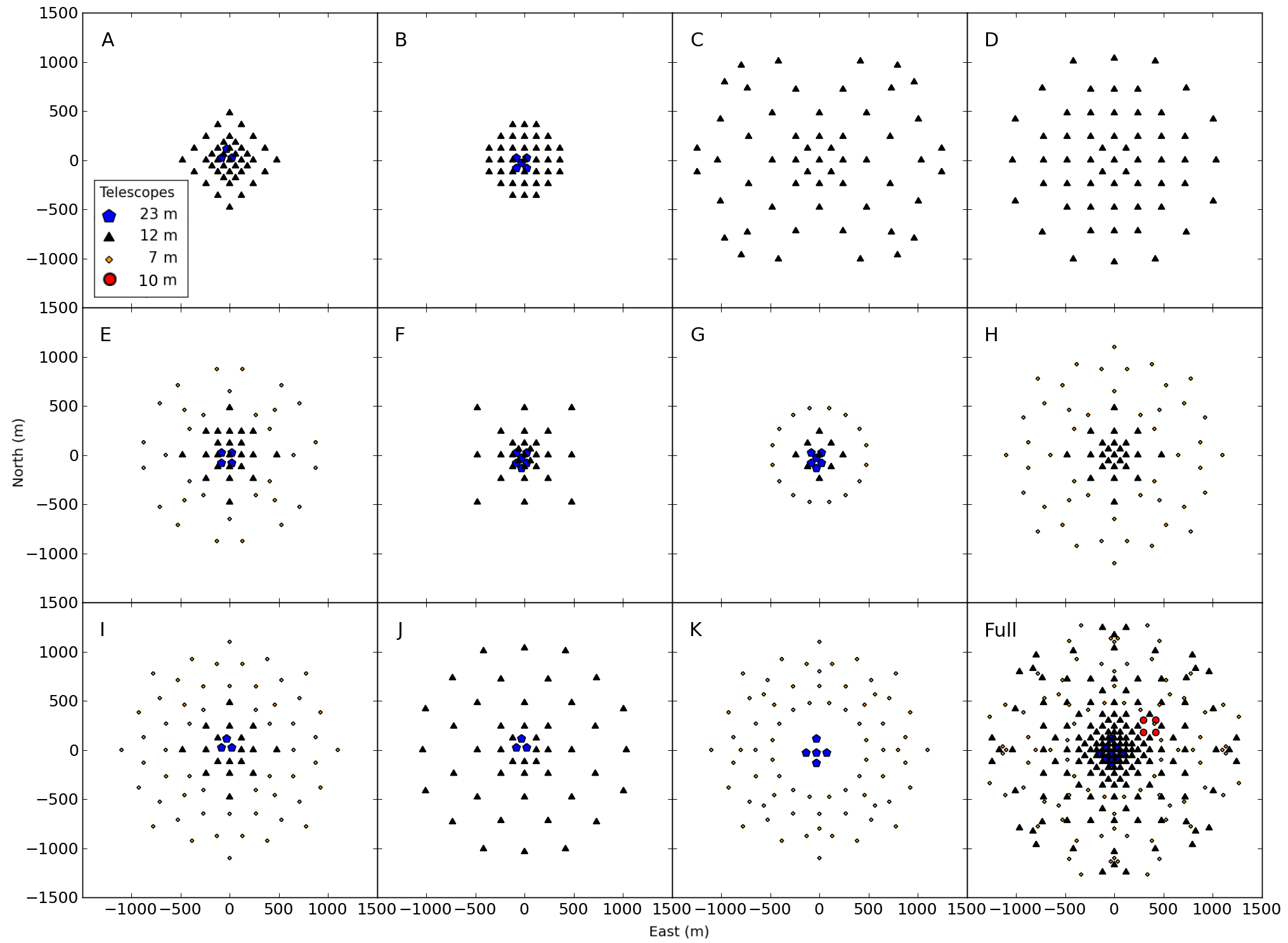} 
\caption{Different telescope array layouts currently being discussed for CTA. Each of the configurations is a subset of the large array shown in the bottom-right corner. For this paper, configurations B, D and I were selected for further study.}
\label{all_configurations}
\end{figure}

A number of candidate array layouts for the CTA are being studied within its design study\cite{benlohr08,ctawebsite,hermann10}, of which a number is shown in Fig.\ \ref{all_configurations}.  Each of these candidate layouts is a subset of an all-encompassing larger array, shown in the bottom-right corner.  Note that several among the configurations are not fundamentally different from each other.   Already from a quick glance, one can roughly divide the layouts into three groups: A, B, F and G, whose telescopes are concentrated in a fairly small area and which lack longer baselines; C, D and J which contain telescopes more sparsely spread out over a large area; and E, H, I and K which contain a larger number of telescopes and provide both long and short baselines.  For this paper, the configurations B, D and I were selected for more detailed study, as ``representatives'' of the three groups.

\begin{figure}[tbh]
\centering
\includegraphics[width=13cm]{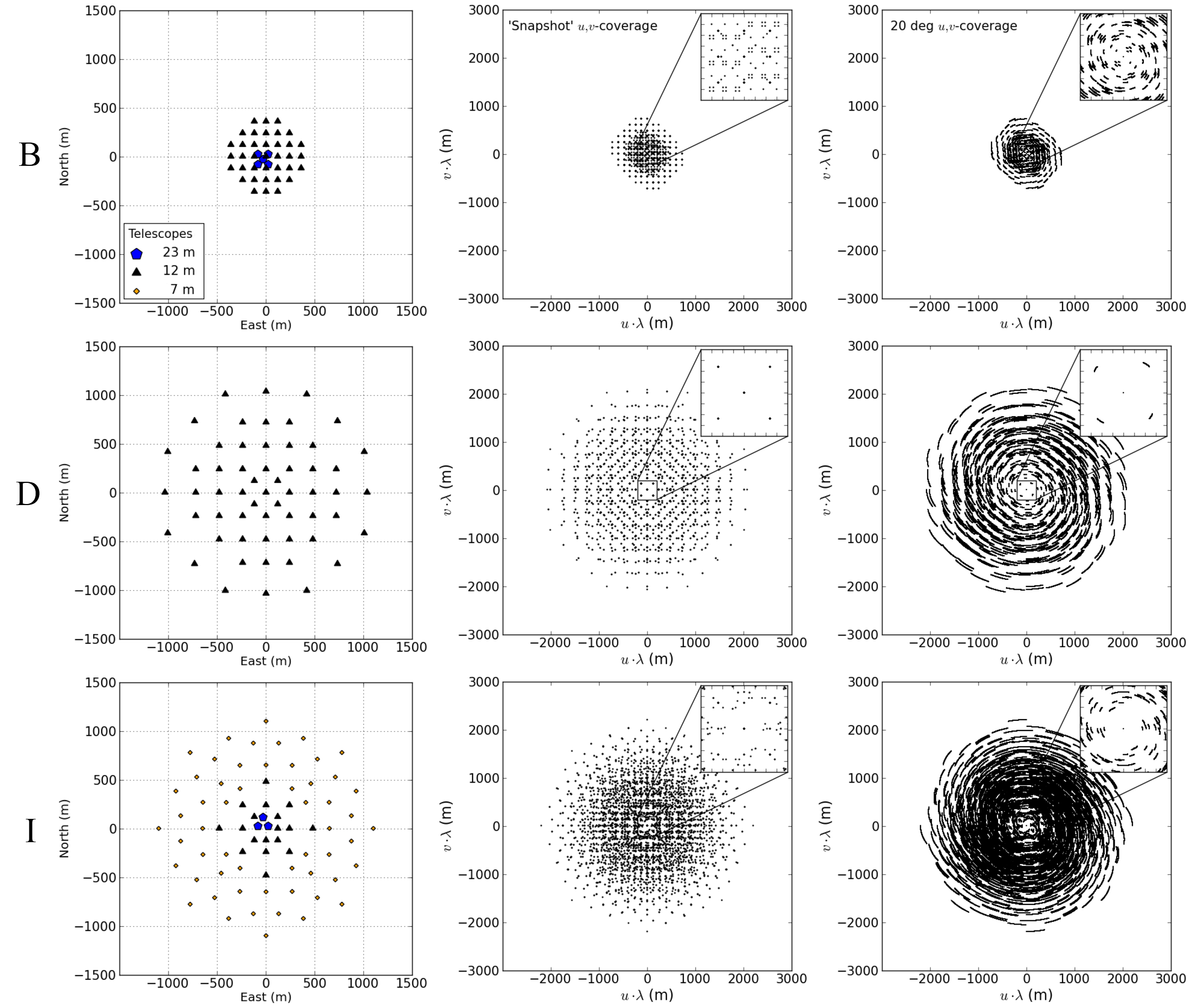} 
\caption{Left column: Telescope placement for the three candidate configurations B, D and I (top to bottom).  Middle column: $(u,v)$-plane coverage at an instant in time for a star in the zenith. Upper right-hand squares show the central 400x400 m area.  Right column: $(u,v)$-plane coverage for a star moving from the zenith through 20 degrees to the west.}
\label{configurations}
\end{figure}

The left column in Fig.\ \ref{configurations} shows the telescope placement of the three examined configurations B, D and I. The middle column shows a ``snapshot'' $(u,v)$-plane coverage of the arrays for a star in the zenith. In the right column, the effect in the $(u,v)$-plane of the star moving 20 degrees from the zenith to the west is shown.

Already from Fig.\ \ref{configurations} and Table \ref{configurationstable} something can be said about the designs if applied to intensity interferometry. Configuration B samples the central parts of the Fourier plane very densely and would provide a field of view that is fully capable of imaging the shapes of almost any stellar object. However, since it lacks longer baselines, it is unable to resolve details smaller than $\sim$0.13 mas.

Configuration D, on the other hand, provides baselines out to 2180 m, permitting studies of very detailed structures, down to around 50 $\mu$as. However, its shortest baseline is 170 m, which means that any structures larger than $\sim$0.6 mas will be lost. This makes it unsuitable for most stars (cf.\ Fig.\ \ref{theta_hist}).  Configuration I seems to provide the best of the two worlds: it has baselines short enough to measure the shapes of the disks of most stellar objects, while still providing very long baselines and a very good resolution.

\section{SIMULATED OBSERVATIONS WITH CHERENKOV TELESCOPE ARRAYS}

An intensity interferometer using two photon-counting detectors $A$ and $B$ and a digital correlator will measure the correlation of Eq.\ \eqref{intcorr} in a discrete form:
\begin{equation}
\label{disccorr}
g^{(2)} = \frac{N_{AB}}{N_A N_B} N,
\end{equation}
where $N_A$ and $N_B$ are the number of photons detected in $A$ and $B$ respectively, $N_{AB}$ is the number of joint detections and $N$ is the number of sampled time intervals. In the simulations presented below, each measurement was simulated by generating random numbers $N_A$, $N_B$ and $N_{AB}$ and inserting into Eq.\ \eqref{disccorr}. These will be Poisson distributed random variables\footnote{In reality, the measurement time is always long enough for the Poisson distributions to be adequately approximated as normal distributions.} with mean values $\lambda_A = P_A\cdot N$, $\lambda_B = P_B\cdot N$ and $\lambda_{AB} = P_{AB}\cdot N$. Here, $P_A$ and $P_B$ are the probabilities of detecting a photon in $A$ and $B$ respectively within a small time interval $\Delta t$ and $P_{AB}$ is the probability of a joint detection within $\Delta t$.

\begin{figure}[h!]
\centering
    \subfigure[Source image]{\label{fig:binarysim-a}\includegraphics[width=5.5cm]{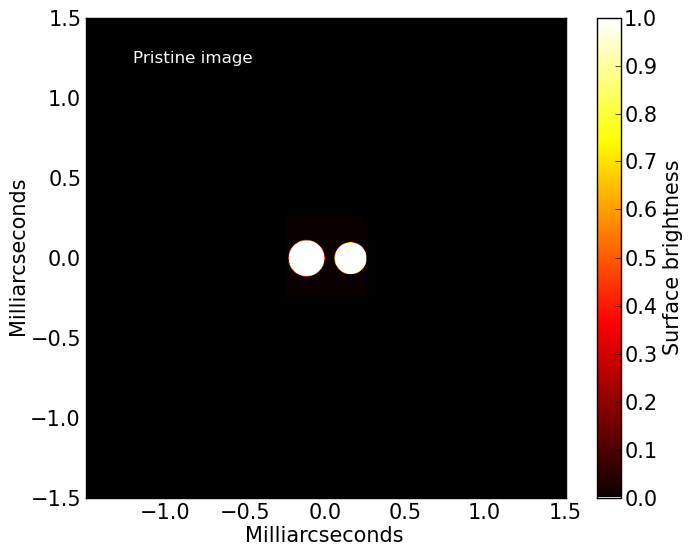}}
        \subfigure[Fourier magnitude]{\label{fig:binarysim-b}\includegraphics[width=6cm]{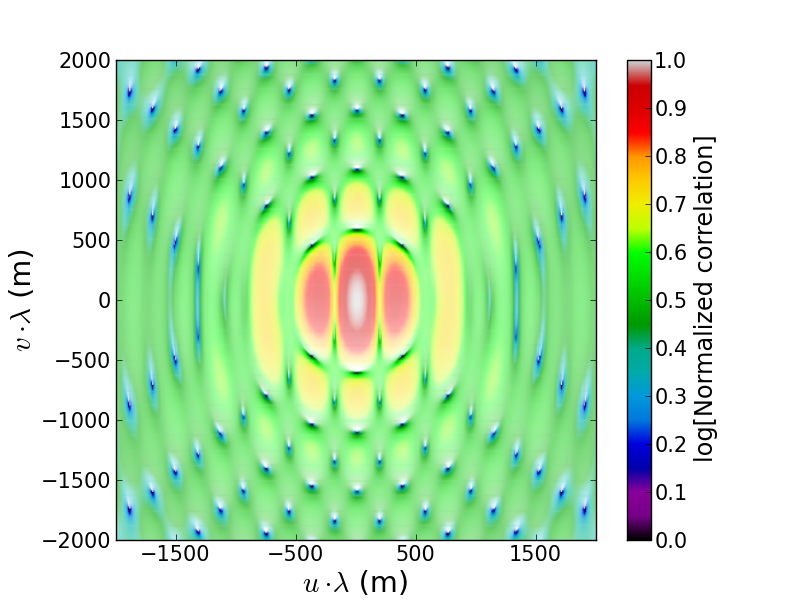}}
  \caption{Test image of a binary star used for simulations and its (logarithmized) Fourier magnitude. }
  \label{fig:binarysim}
\end{figure}

These probabilites can be written out in terms of variables depending only on the instrumentation and the target of study (e.g. Ref.\ \citenum{mandel95}):
\begin{align}
P_A &= \alpha_A \langle I_A \rangle \Delta t \\
P_B &= \alpha_B \langle I_B \rangle \Delta t \\
P_{AB} &= P_A P_B + \alpha_A \alpha_B \langle I_A \rangle \langle I_B \rangle |\gamma_{AB}|^2 \frac{\tau_c}{\Delta t} \Delta t^2 
\end{align}
Here $\alpha$ denotes the quantum efficiency\footnote{The quantum efficiency of a detector is essentially the ratio of detected photons to incoming photons.} of the detectors, $\langle I \rangle$ is the mean light intensity, $\tau_c$ is the coherence time of the light (determined by the wavelength and optical passband) and $\gamma_{AB}$ is the degree of optical coherence (proportional to the Fourier transform of the target).  Such simulations were carried out for the various candidate  configurations using the binary star in Fig.\ \ref{fig:binarysim} as the source to be observed. 

\begin{figure}[h!]
\centering
\includegraphics[width=13cm]{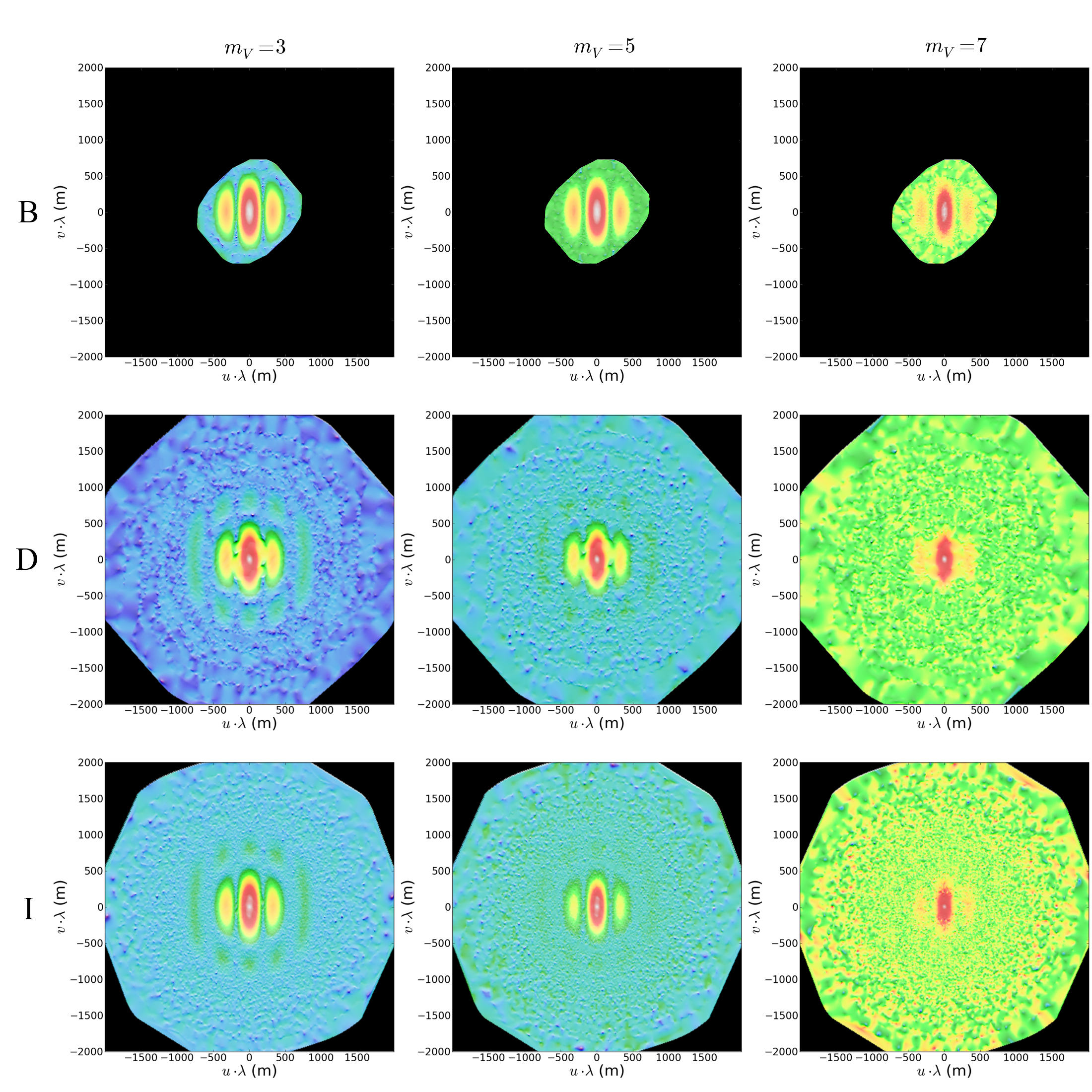} 
\caption{Simulated observations of the binary star in Fig.\ \ref{fig:binarysim} with configuration B (top row), D (middle) and I (bottom).  Vertical columns are for differently bright stars; left is for $m_V=3$; center for $m_V=5$, and $m_V =7$ to the right. }
\label{magnitudes}
\end{figure}

\subsection{Varying target magnitude}

Figure \ref{magnitudes} shows the results of varying the brightness of the target. The simulations started with the star in zenith and then moving 50 degrees to the west in steps of 5 degrees, with an integration time of 2 hours at each time step (thus, such an observation would require several nights of integration). The observational wavelength was 500 nm and the detectors were assumed to have a time resolution of 1 ns and a quantum efficiency of 70 \%. 

The data from the simulations are shown in the Fourier plane.  To better illustrate the locations of low-amplitude structures and of noise, the data were logarithmized and normalized after first performing a linear interpolation between the ``observed'' points, in order to obtain the Fourier magnitude over a regular grid.  A 3D shading was then applied to the data to further highlight the locations of structures (the absolute values of the measured correlations are not of any real significance in this context).

These simulations are consistent with earlier results in that the practical limiting magnitude seems to lie somewhere around $m_V = 6-7$.  Configuration B performs well in reproducing the three central lobes of Fig.~\ref{fig:binarysim-b}, but it lacks the longer baselines needed for higher-frequency components. 

Configuration D, on the other hand, measures the higher-frequency components very well but even for a target with such a small angular extent, the central parts of the Fourier plane -- which determine the overall shape of the object -- are distorted due to poor sampling at short baselines.  Configuration I samples the whole Fourier plane well, but its smaller-size telescopes (see Fig.\ \ref{configurations}) makes it a bit more sensitive to target magnitude.

\begin{figure}[h!]
\centering
\includegraphics[width=13cm]{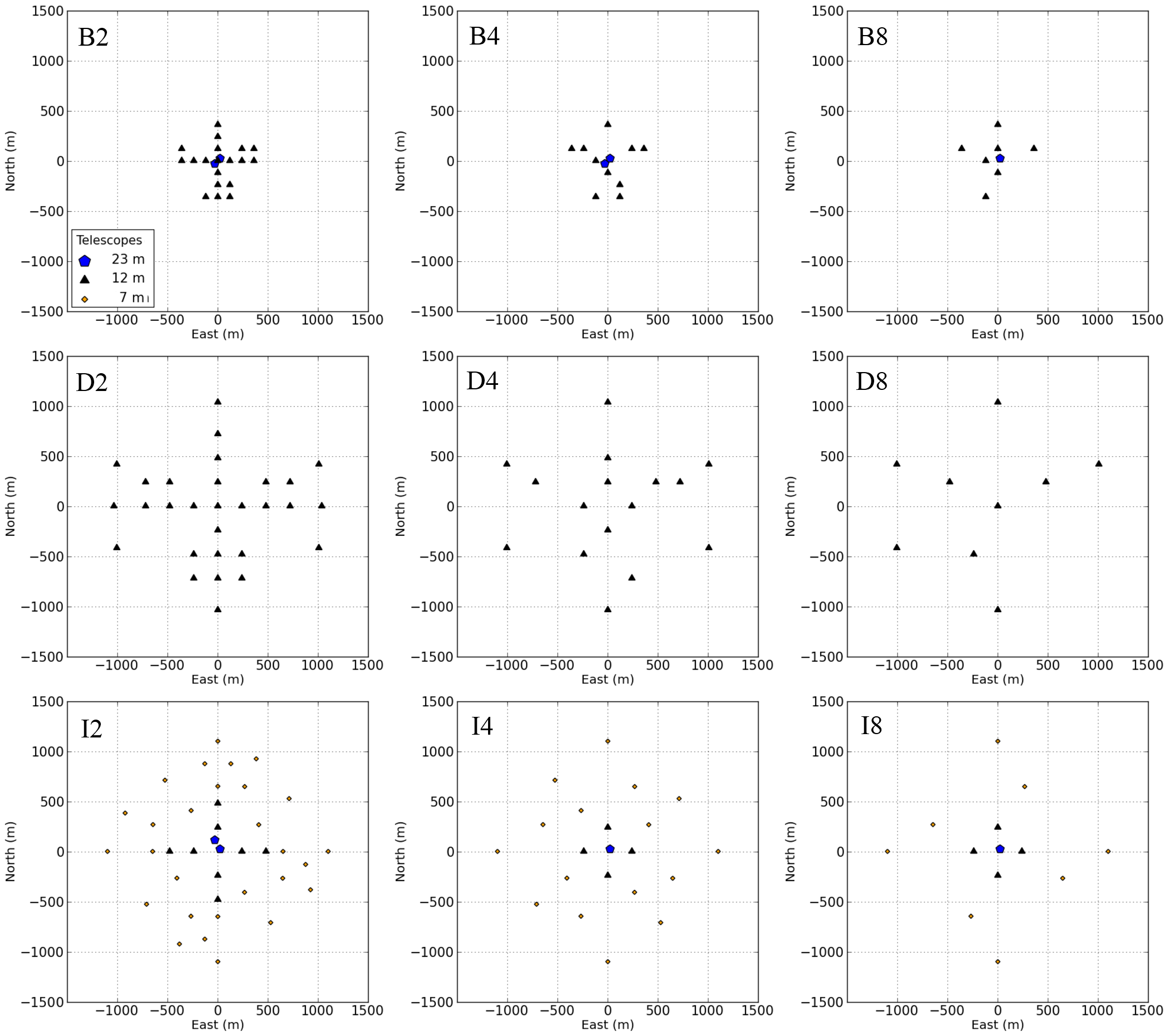} 
\caption{Subsets of the candidate configurations (B, D, I from top to bottom). In the leftmost column half the telescopes of the superset configuration were selected in a pseudo-random fashion. In the middle column, one in four telescopes was selected, and in the rightmost column one in eight.}
\label{subsets_configs}
\end{figure}

\subsection{Observations with subsets of the configurations}
It is interesting to examine the performance of also various smaller subsets of the telescope arrays.  It is not known whether some modifications or special equipment might be required at each telescope to be used for intensity interferometry, nor whether the full array will always be available.  Observations using only a subset of the telescopes may thus represent a realistic mode of operation, at least initially.

For each of the configurations B, D and I, three subset configurations were generated, as shown in Fig.\ \ref{subsets_configs}. The configurations shown in the leftmost column, designated B2, D2 and I2, were obtained by selecting half of the full set of telescopes in a semi-random manner (attempting to preserve the overall ``shape'' of the array). In the middle column, one in four telescopes was retained (these are referred to as B4, D4, and I4) and in the rightmost column only one telescope in eight was kept (B8, D8, and I8). 

\begin{figure}[tbh]
\centering
\includegraphics[width=13cm]{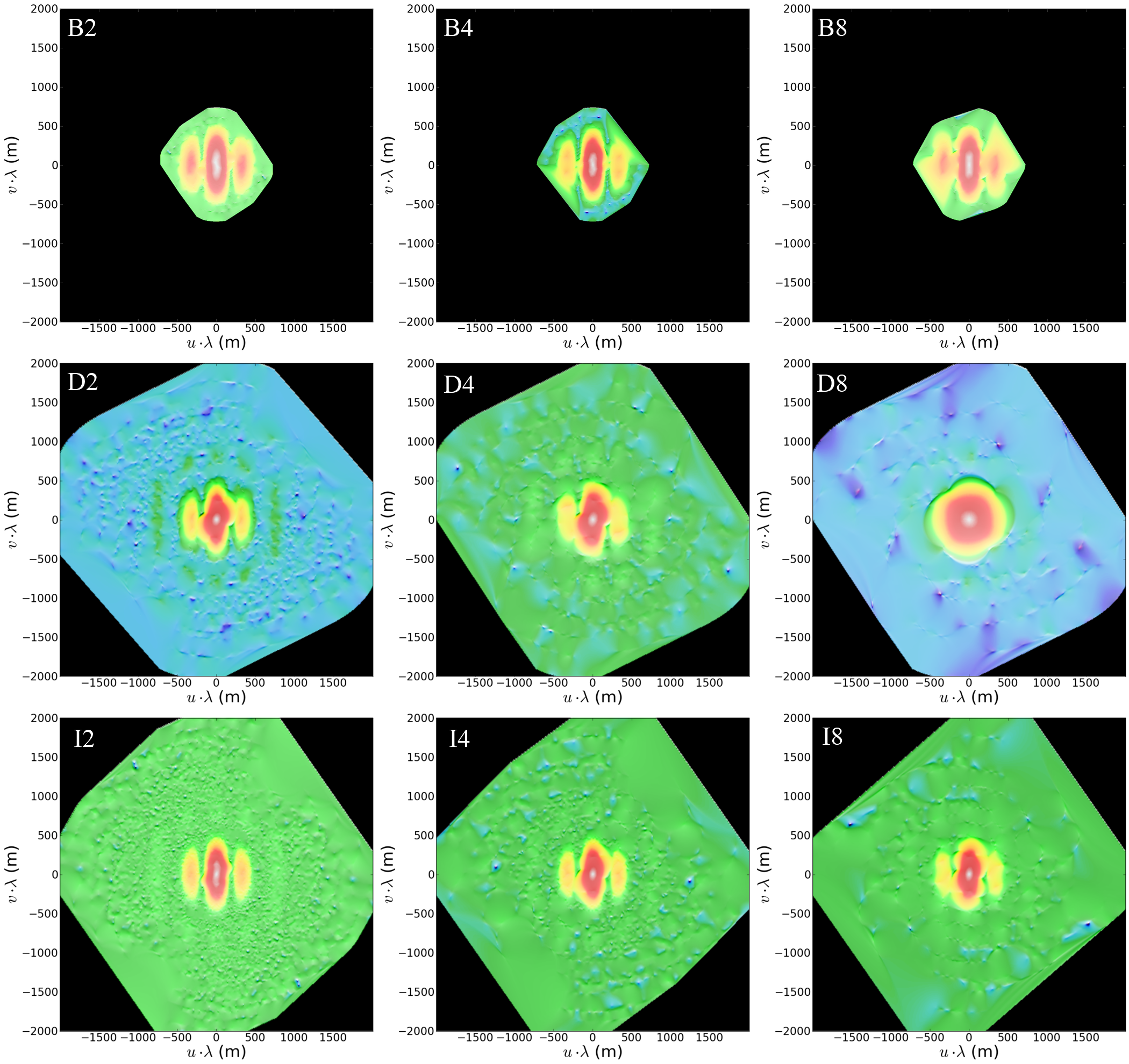} 
\caption{Simulated observations of the binary star in Fig.\ \ref{fig:binarysim} with subsets of configuration B (top row), D (middle row) and I (bottom row).  The left column is for subsets containing half of the telescopes; the center for a quarter, and the rightmost for one eighth (see Fig.\ \ref {subsets_configs}). }
\label{subsets}
\end{figure}

Figure \ref{subsets} shows the output from simulations of the binary star in Fig.\ \ref{fig:binarysim-a} using these subsets. The simulation parameters were the same as in Fig.\ \ref{magnitudes} except that the magnitude of the star was now fixed to $m_V$ =5 and the integration time was increased to 10 hours per timestep in order to depress measurement noise and thus highlight sampling effects for the various configurations.

It is obvious that more telescopes are better in terms of Fourier plane sampling, and too few telescopes can ruin the results, as illustrated by the observation with D8. It should be noted, however, that no attempt was made to optimize the selection of telescopes in order to improve the sampling.  Once the design for CTA is finalized, this could be one of the first problems to examine in order to determine which telescopes to equip first for intensity interferometry.

\subsection{Image reconstruction}

The discussion above has only dealt with Fourier magnitudes, which is what is directly measured by an intensity interferometer. Fourier magnitudes alone can be used to fit model parameters such as stellar diameters, binary separations, circumstellar disk thicknesses, etc. As is shown in Ref.\ \citenum{nunez10}, model-independent image reconstruction from the Fourier magnitude alone is indeed possible (though with some limitations).

\begin{figure}[ht]
\centering
\includegraphics[width=9cm]{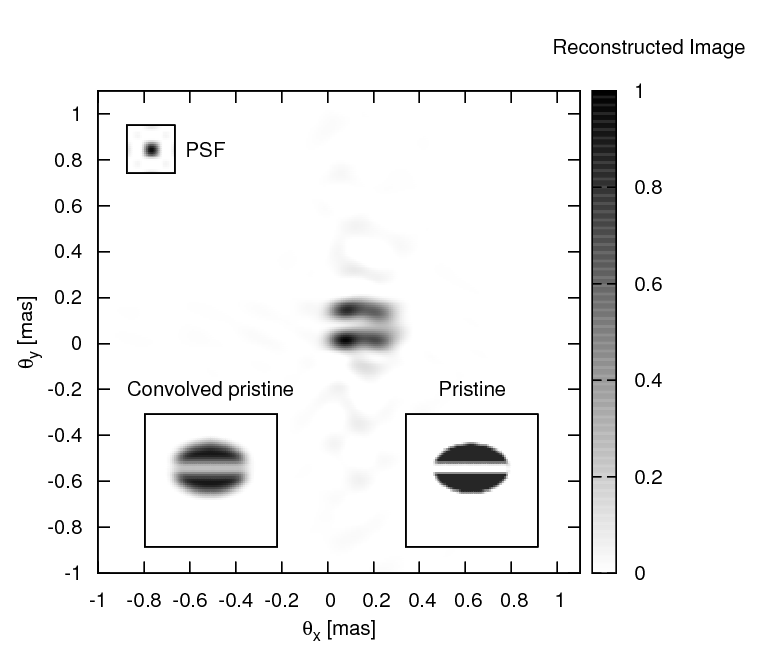}
\label{thick_disk_cta}
\caption{Image reconstructed from simulated data corresponding to 50 hours of observation of a magnitude 6 rotationally flattened star with an obscuring disk. The simulation was done using a telescope array qualitatively similar to those discussed in this paper (see Ref.\ \citenum{nunez10} for details). The image in the bottom-right corner was used as input for the simulation. Also shown is the pristine image convolved with the point-spread-function of the array. The difference in sharpness between this and the pristine image is due to the inability of a finite-sized array to measure all the higher-frequency Fourier components. The convolved pristine image is thus the best image that is theoretically possible to reconstruct.}
\label{thick_disk_cta}
\end{figure}

Figure \ref{thick_disk_cta} shows a proof-of-concept example of an image reconstructed from simulated data using one of the candidate configurations. However, image reconstruction is an active area of research and many practical issues remain to be improved. Thus, at this stage, it is not yet possible to tell which imperfections are due to fundamental limitations of the algorithms used or simply due to practical issues that can be solved. There is thus a risk that a comparison between reconstructed images from the different candidate arrays would be misleading, which is why the focus here is instead on the algorithm-independent Fourier magnitudes.

\section{Conclusions}

A facility such as the CTA will most likely be able to produce excellent scientific data if used as an intensity interferometer, providing measurements of spatial structures on scales down to a few tens of $\mu$as -- orders of magnitude better than what is possible today in visible light.

The candidate configurations examined here -- while not optimized for intensity interferometry -- all provide very dense sampling of the $(u,v)$-plane due to the sheer number of telescopes.  Apart from the simulations described above, a number of other tests were carried out to determine effects from varying the declination of the star and the geographic orientation of the telescope array. It was found that the very large number of telescopes make the effects of any such variations almost negligible.

However, it must be noted that an array such as configuration D will be severely crippled by its lack of short baselines, essentially limiting such a facility to only studying stars smaller than $\sim$0.5 mas.  Also, to improve coverage of the $(u,v)$-plane and to ensure a wider variety of baselines (enabling better image reconstruction), an exact east-west orientation should be avoided for the grid of many smaller telescopes (since stars rise in the east, and move towards west, the projection of any east-west baseline will vary only in amount, not orientation).  For the fewer large telescopes, repetitive geometric patterns should be avoided, in order to avoid redundancy in baselines.  To cover astrophysically relevant dimensions between 0.1--3 milliarcseconds, baselines between pairs of telescopes should as far as possible cover the whole interval 30--2000 m.

\acknowledgments
The work at Lund Observatory is supported by the Swedish Research Council and The Royal Physiographic Society in Lund. S. LeBohec acknowledges support from grants SGER \#0808636 of the National Science Foundation.


\bibliography{report}   
\bibliographystyle{spiebib}   

\end{document}